# A Block-Ring connected Topology of Parameterized Quantum Circuits

Wenjie Liu [1,2*], Qingshan Wu[1], Ying Zha[1]

1. School of Software, Nanjing University of Information Science and Technology, No. 219 Ningliu Road, Nanjing, 210044, Jiangsu, China
2. Engineering Research Center of Digital Forensics, Ministry of Education, No. 219 Ningliu Road, Nanjing, 210044, Jiangsu, China.

E-mail: wenjieliu@nuist.edu.cn

**Abstract**

It is essential to select efficient topology of parameterized quantum circuits (PQCs) in variational quantum algorithms (VQAs). However, there are problems in current circuits, i.e. optimization difficulties caused by too many parameters or performance is hard to guarantee. How to reduce the number of parameters (number of single-qubit rotation gates and 2-qubit gates) in PQCs without reducing the performance has become a new challenge. To solve this problem, we propose a novel topology, called Block-Ring (BR) topology, to construct the PQCs. This topology allocate all qubits to several blocks, all-to-all mode is adopt inside each block and ring mode is applied to connect different blocks. Compared with the pure all-to-all topology circuits which own the best power, BR topology have similar performance and the number of parameters and 2-qubit gate reduced from $O(n^2)$ to $O(mn)$, $m$ is a hyperparameter set by ourselves. Besides, we compared BR topology with other topology circuits in terms of expressibility and entangling capability. Considering the effects of different 2-qubit gates on circuits, we also make a distinction between controlled X-rotation gates and controlled Z-rotation gates. Finally, the 1- and 2-layer configurations of PQCs are taken into consideration as well, which shows the BR's performance improvement in the condition of multilayer circuits.

## 1. Introduction

With the further development of quantum algorithms and devices, we have reached the era of "Noisy Intermediate-Scale Quantum" (NISQ)(Preskill 2018). The disadvantages of NISQ devices are obvious, they only have 50 to 100 qubits, support no more than 1000 quantum gates, and the error-corrected is unavailable. Although these NISQ devices currently can not complete large-scale quantum computation or realize great commercial value, we can still use them to realize or explore the more potential of quantum computers with the supply of classical computers. Under the guidance of this strategy, a plenty of variational quantum algorithms (VQAs) have been proposed. For example, variational quantum eigensolver (VQE) is applied to compute the ground states of molecular systems(Hastings, Wecker et al. 2015, McClean, Romero et al. 2016, Cao, Romero et al. 2019). These VQAs explore a new approach to solve some problems about linear algebraic.

Other VQAs, such as quantum approximate optimization algorithm (Farhi, Goldstone et al. 2014) solves the problem of max-cut and number partitioning; quantum variational autoencoder (Romero, Olson et al. 2017), quantum machine learning (QML) for classification(Farhi and Neven 2018, Wilson, Otterbach et al. 2018, Havlíček, Córcoles et al. 2019, Schuld, Bocharov et al. 2020), Quantum generative adversarial networks (QGAN)(Dallaire-Demers and Killoran 2018, Lloyd and Weedbrook 2018, Zhu, Linke et al. 2019) focus on solving various problems by using quantum neural networks (QNN) based on PQCs. If we want to realize the algorithms or the neural networks mentioned above, constructing parameterized quantum circuits is a core step that cannot be ignored.

Although PQCs play an important role in VQAs, the research on PQCs is still in its infancy. At present, there are few quantum circuit topologies we can adopt in HQC. How to choose a suitable PQCS is extremely important for researchers. In (Sim, Johnson et al. 2019), expressibility and entangling capability as a measure of the performance of a PQCS were proposed, 19 specific circuits based on three topologies are summarized and compared. These works may provide a obvious guide for researchers. However, with the expansion of the scale of the problem, the number of qubits, the number of parameters, layers and the depth of of PQCS also increase sharply. This also brings many problems, such as barren plateaus(McClean, Boixo et al. 2018, Pesah, Cerezo et al. 2021) will make the loss function converge slow or convergence is unreachable. In this case, the selection of efficient shallow-circuit is particularly important.

However, current topologies for constructing PQCs has many disadvantages, i.e. optimization difficulties caused by too many parameters or performance is hard to guarantee. To solve these problems, we propose a novel topology, called Block-Ring (BR) topology. BR topology can save number of parameters without reducing the performance.

The structure of this paper is as follows. Before presenting our topology, Three topologies in PQCs, the expressibility, entangling capability and their specific measurement methods are introduced briefly in Section 2. In Section 3, we propose a novel topology of PQCs called BR topology, the whole circuit consists of several qubit-blocks: inside the blocks, qubits are connected follow a all-to-all mode, ring connected topology is adopted between blocks.

## 2. Preliminaries

In this section, we firstly introduce the definition of parameterized quantum circuits. Then, two descriptors named expressibility and entangling capability defined in (Sim, Johnson et al. 2019) are investigated and the methods of calculating these two descriptors are also given.

### 2.1 Three topologies in PQCs

Parameterized quantum circuits (PQCs) are also sometimes referred to as quantum neural networks (QNN) because the application of PQCs for machine learning problems are similar to classical neural networks (NN). PQCs are usually defined as a unitary operation with variable parameter on $n$ qubits, $U_\theta$. This operation will transform an initial state $\phi_0$, often set to $|0\rangle^n$, into a parameterized quantum state:

$$|\psi_\theta\rangle = U_\theta |\phi_0\rangle, \#(1)$$

where $\theta$ is a vector of parameters. A general PQC is shown in Fig. 1

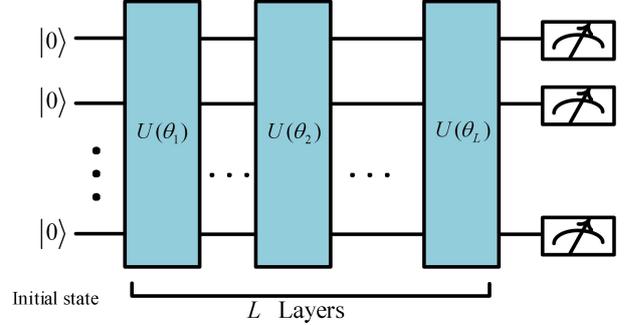

Fig. 1 A general circuit of PQCs.

One operation by applying $U_\theta$ on qubits can be regarded as a circuit template containing single-qubit gates and 2-qubit gates. Single-qubit gates contains *Rz*, *Ry*, *Rx*, and 2-qubit gates usually include *CRx*, *CRz*. In order to provide more variability, this circuit template can be repeated $L$ times to construct a deeper circuit with more parameters.

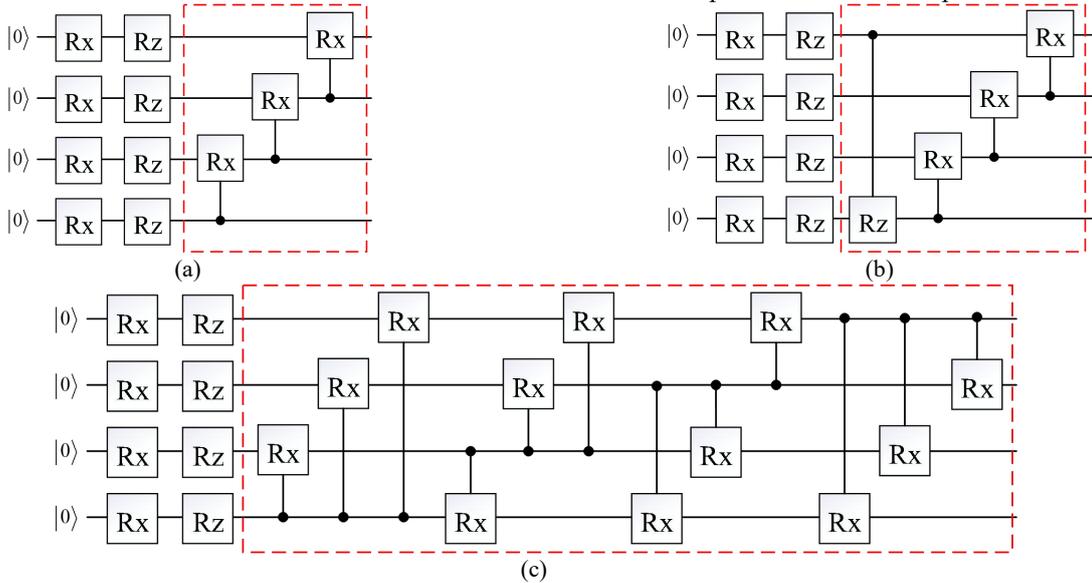

Fig. 2 Three topologies of 4-qubit circuit. (a) the line topology, (b) the ring topology, and (c) the all-to-all topology.

Different PQCs are often constructed by different topology. As shown in Fig. 2, there are three common topologies for buiding a PQC, line topology, ring topology and all-to-all topology. Ring topology adds a regression control based on linear topology. In practical applications, the three topologies in Fig. 2 are often regarded as single-layer templates, which can be used repeatly for construct a deep QNN.

## 2.2 Expressibility and Entangling capability

(a) Expressibility (*Expr*) delegates the ability to generate (pure) states of a circuit. Comparing the distribution of state generated by a PQC with random sampling parameter to the uniform distribution of state tend to be a approach for quantifying the expressibility of a circuit. In (Sim, Johnson et al. 2019), the gap between the quantum state generated by PQC and the quantum state of uniform distribution can be written as:

$$A = \int_{Haar} (|\psi\rangle\langle\psi|)^{\otimes t} d\psi - \int_{\Theta} (|\varphi_\theta\rangle\langle\varphi_\theta|)^{\otimes t} d\theta . \#(2)$$

The *Expr* of a PQC can be calculated by Kullback-Leibler (KL) divergence(Kullback and Leibler 1951):

$$Expr = D_{KL}\left(\hat{P}_{PQC}(F;\theta) || P_{Haar}(F)\right). \#(3)$$

The lower score obtained by this method means the PQC may be a more expressible circuit.

(b) Entangling capability (*Ent*) intends to capture the capability to generate highly entangled state. (Sim, Johnson et al. 2019) proposed the computation of entangling capability can be completed by Meyer-Wallach (MW) entanglement measure (Meyer and Wallach 2002).

The generalized distance $D$ of two states $|u\rangle = \sum u_i |i\rangle$, $|v\rangle = \sum v_i |i\rangle$ can be described as:

$$D(|u\rangle, |v\rangle) = \frac{1}{2}\sum_{i,j}|u_i v_j - u_j v_i|^2 . \#(4)$$

Then, the MW entanglement measure, namely $Q$, is defined as follows,

$$Q(|\psi\rangle) \equiv \frac{4}{n}\sum_{j=1}^{n} D\left(\iota_j(0)|\psi\rangle, \iota_j(1)|\psi\rangle\right). \#(5)$$

where

$$\iota_j(b)|b_1 \ldots b_n\rangle = \delta_{bb_j}|b_1 \ldots \hat{b}_j \ldots b_n\rangle, \#(6)$$

$$\delta_{bb_j} = \begin{cases} 1, b = b_j \\ 0, b \neq b_j \end{cases}, \#(7)$$

$b_j \in \{0,1\}$, and $\hat{b}_j$ represents the absence of the *j*-th qubit.

For a PQC, the entangling capability can be estimated with the following formula,

$$Ent = \frac{1}{|S|}\sum_{\theta_i \in S} Q(|\psi_{\theta_i}\rangle) \#(8)$$

where $S = \{\theta_i\}$, $\theta_i$ is the *i*-th parameter vectors of PQC.

## 3. A proposed Block-Ring topology of PQCs

In this section, we propose a novel topology called Block-Ring topology . We first describe the characteristic of BR topology and the way of connecting qubits. Then, algorithm about how to build BR topology circuit is given. Finally, 8-qubit BR topology circuits are given as a example.

### 3.1 A proposed topology

An efficient PQC plays a very important role in HQC algorithm. How to saving the number of parameters without reducing the performance of PQC has become a new challenge. BR topology is proposed to solve this problem. BR topology can be considered as the synthesis of ring mode and all-to-all mode.

In the proposed BR topology, $n$ qubits are divided into $t$ blocks on average, $B = \{B_1, B_2, \ldots, B_t\}$, where $t = n/m$, $b_i$ is the *i*-th block ($i = 1, 2, \ldots, t$), $m$ is the number of qubits in a block. Inside each block, the all-to-all mode is applied to connect qubits $\{b_{i1}, b_{i2}, \ldots, b_{im}\}$, where $b_{ij}$ represents the *j*-th qubit ($j = 1, 2, \ldots, m$) of the *i*-th block. As shown in Fig. 3, all vertices in the block form a fully connected graph, i.e. there is an entanglement operation between any two different qubits $\langle b_{iq}, b_{ip}\rangle = 1$, where $q \neq p$. In addition, all blocks are connected in a ring mode. Specifically, the qubits between adjacent blocks form a one-to-one mapping relationship, two qubits at the same index in adjacent blocks are connected, where $\langle b_{hj}, b_{\hat{h}j}\rangle = 1$, $h = 1, 2, \ldots, t$, $\hat{h} = mod(h + 1, t)$.

Fig. 4 shows an example of 9-qubits BR topology. In this example, there are 3 blocks, and each block owns 3 qubits. Inside each block, these 3 qubits (represented with 3 different shapes in Fig. 4) are connected with each other. Between two adjacent blocks, the qubits with the same shape are connected, and finally form a "ring". In Fig. 5, we give the whole 9-qubit BR topology circuit, the operations in the green dashed box make up the ring topology between blocks, the operations in red dashed box means all-to-all connection inside a block.

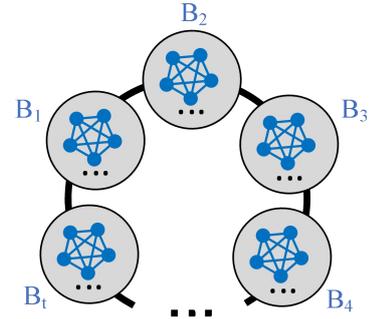

Fig. 3 BR topology with *t* blocks. Each dot represents a qubit, *m* qubits form a block, and *t* blocks are connected in a ring.

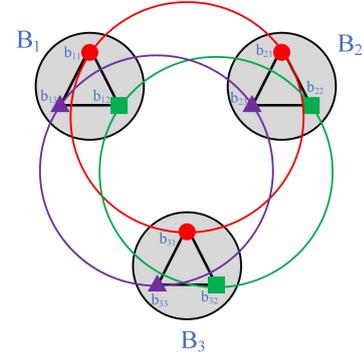

Fig. 4 Example of 9-qubit BR topology with 3 blocks. The qubits with the same shape (triangles, circles, squares) are connected to each other.

### 3.2 Algorithm for constructing BR topology circuit

Algorithm 1 gives the process of constructing the circuit in detail. $CR(qubit1, qubit2)$ is 2-qubit operation, *qubit1* is control qubit and *qubit2* is controlled qubit. In some previous papers, researchers often used $CRx$ or $CRz$ gate to implemnt entangled state. In this paper, all circuits are constructed by these two kinds of 2-qubit gate and compared separately.

There are 2 extreme cases in **Algorithm** 1, if $m$ equals 1 or $n$, we will get a circuit like ring topology or all-to-all topology. In the event of $n$ is not a prime number, according to **Algorithm** 1, we can build circuits with arbitrary number of qubits.

**Algorithm** 1. Constructing BR topology circuit

**Input**: Number of qubits $n$, number of qubits $m$ in a block.
**Output**: BR topology circuit.
1: add $Rx, Rz$ gates on $n$ qubits;
2: set $i = 0$;
3: **while**$(i < \frac{n}{m} - 1)$:
4:     set $j = 0$;
5:     **while**$(j < m)$:
6:         $CR(qubits[m * i + j], qubits[m * (i + 1) + j])$;
7:         $j = j + 1$
8:     $i = i + 1$
9: set $i = 0$;
10: **while** $\left(i < \frac{n}{m}\right)$:
11:     All-to-all mode is applied on $block[i]$;
12:     $i = i + 1$
13: set $j = 0$
14: **while**$(j < m)$:
15:     $CR(qubits[n - m + j], qubits[j])$;
16:     $j = j + 1$
17: add $Rx, Rz$ gates on n $n$ qubits;
18: **Return** circuit;

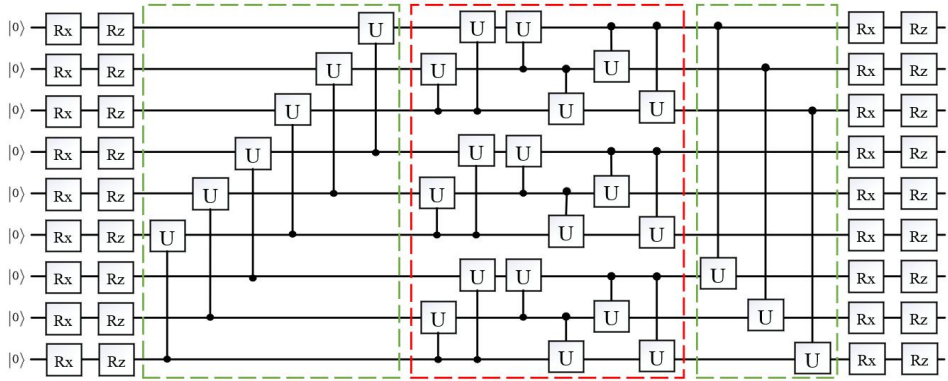

Fig. 5 BR topology of 9-qubit circuit. Operation U can be chosen from $\{Rx, Rz\}$.

## 4. Experiment

In order to verify the superiority of our BR topology, we selected several classes of 8-qubit circuits (Fig. 6-9) that perform better in (Sim, Johnson et al. 2019) as comparison in terms of "*Expr*" and "*Ent*". Circuit 1-2 in Fig. 6 are built by line topology; Circuit 3-4 in Fig. 7 belong to ring topology; Circuit 5-6 in Fig. 8 are varieties of circuit 3-4; Circuit 7-8 in Fig. 9 are constructed by all-to-all topology. Circuits in Fig. 10 are constructed by our BR topology. To explore the effect of increasing the number of circuit layers on the expressibility and entangling ability, 2-layer configuration of circuits is considered and tested.

### 4.1 Experiment setting

The experiment environment is as follows: CPU is Intel Xeon(R) Gold 5220R and memory is 256GB. IBMQ is chosen as quantum simulator. And here is our experiment setup: For each experiment, we make 20480 measurements for each circuit to construct its distribution, using a bin size of 75. After 10000 experiments, we counted the distribution samples and calculated the KL divergence value.

### 4.2 Result

Fig. 11 and Fig. 12 present the final experiment results. Red dots represent the circuits with single layer. Blue dots represent the circuits with 2 layers. In Fig. 11, we can find that circuit 7 and circuit 9 have better expressibility in the condition of single layer, but the expressibility of circuit 9 is little lower than that of circuit 7. This phenomenon is normal because the number of parameters of circuit 9 is much less than that of circuit 7. Once the circuit layer repeats twice, the expressibility of circuit 7 has not increased much, the expressibility of circuit 9 has caught up with that of circuit 7. Although circuit 5 has the highest rise rate on expressibility, its expressibility still lower than circuit 7.

In Fig. 12, circuit 7 always have best entangling capability. The entangling capability of circuit 9 is slightly lower in the condition of single layer, its entangling capability can still catch up with circuit 7 after the circuit layer is repeated twice. It can be considered that the performance of circuit 9 is no worse than that of circuit 7.

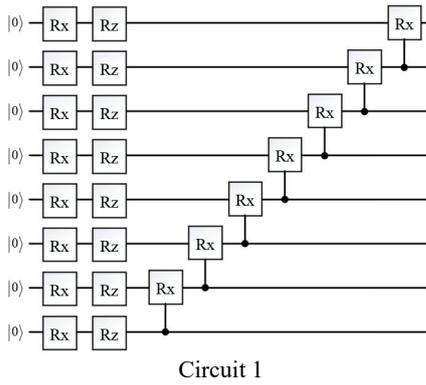
Circuit 1
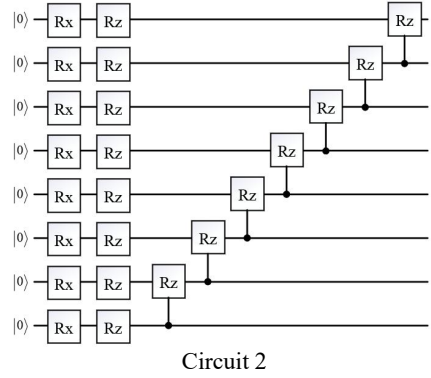
Circuit 2

Fig. 6 8-qubit circuits constructed by line topology.

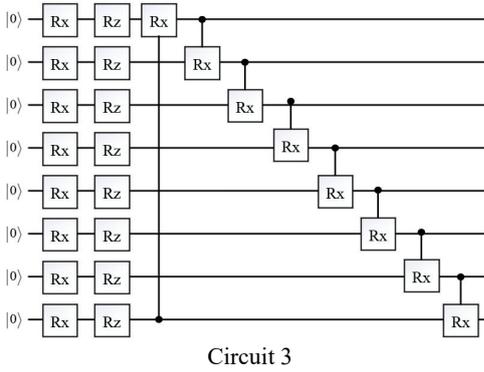
Circuit 3
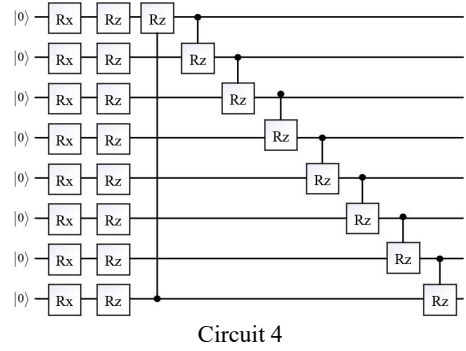
Circuit 4

Fig. 7 8-qubit circuit constructed by ring topology.

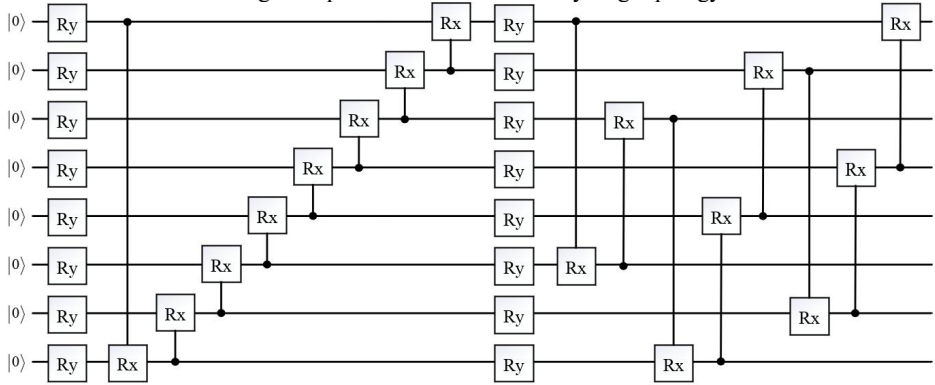
Circuit 5

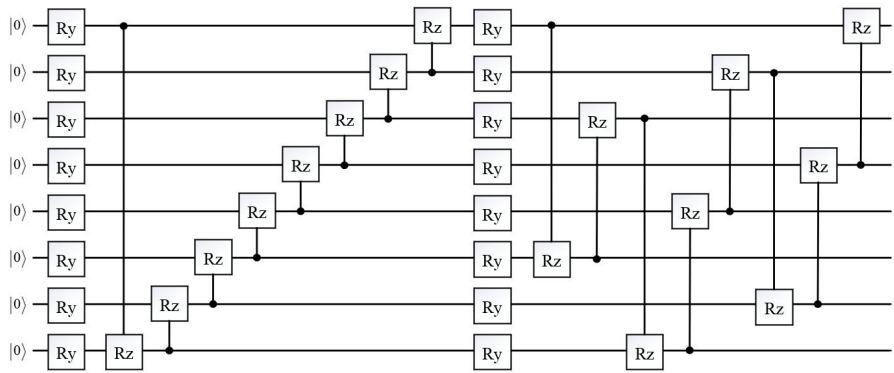
Circuit 6

Fig. 8 8-qubit circuit constructed by ring topology with double layer.

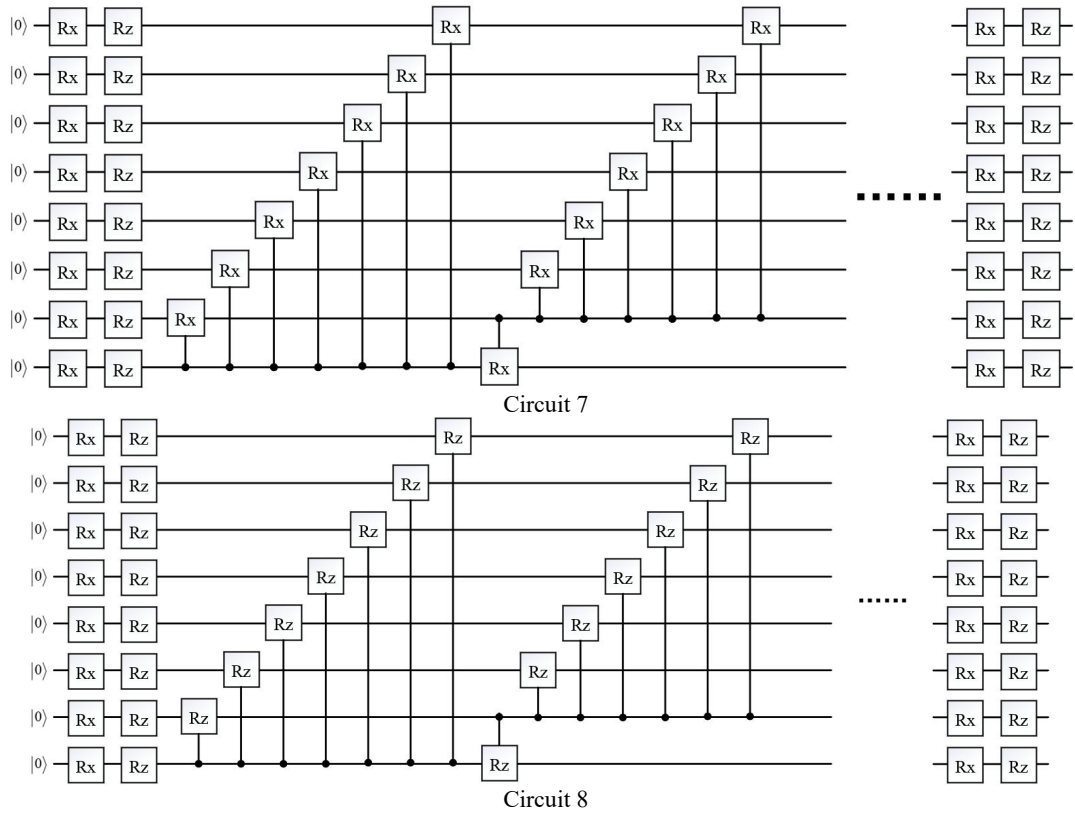

Circuit 7

Circuit 8

Fig. 9 8-qubit circuit constructed by all-to-all topology.

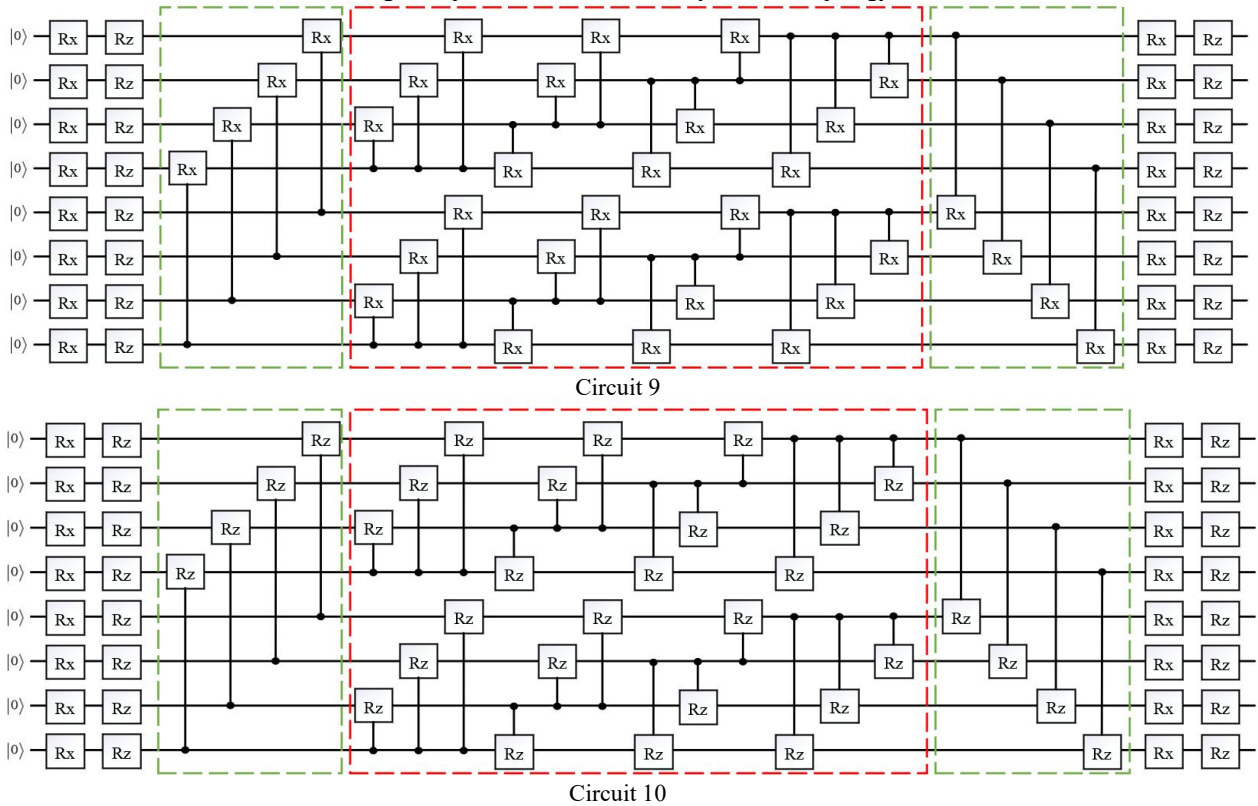

Circuit 9

Circuit 10

Fig. 10 8-qubit circuit constructed by BR topology.

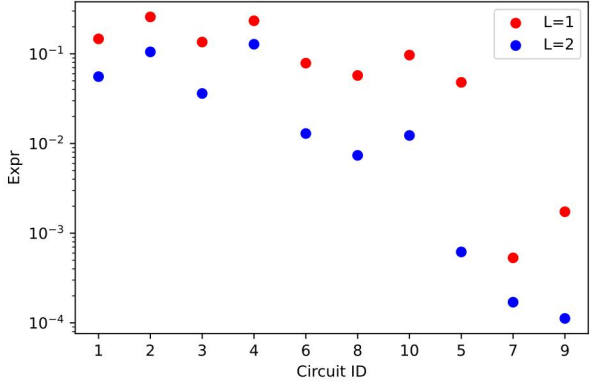

Fig. 11 Expressibility of 10 circuits.

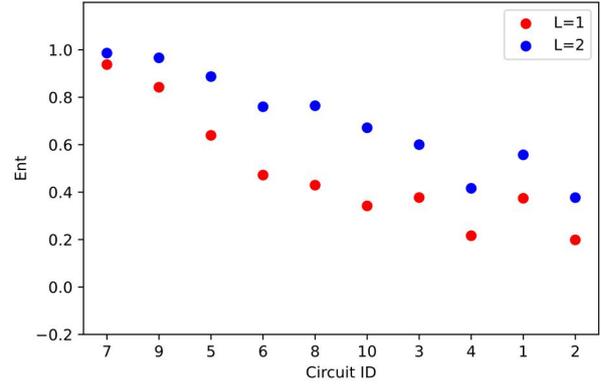

Fig. 12 Entangling capability of 10 circuits.

## 5. Cost analysis

The complexity of quantum circuits depends on the number of parameters, the number of 2-qubit gates, and the circuit depth. In this section, we analyzed the cost of our BR topology and compare it with other two circuits (Circuit 5, 7) in aspects mentioned above.

In BR topology with single layer, the number of 2-qubit gates used in ring mode is $n$, the operations inside single block spend $m(m-1)$ gates. For a whole circuit, the number of 2-qubit gates is $\left(n + \frac{n}{m} * m(m-1)\right)L = mnL$. $L$ is the number of layer. Considering the parameters in single-qubit gate, the total number of parameters is $(m+4)nL$. Furthermore, the circuit depth of the BR topology must be highlighted. In a layer, since the operations in ring mode and all-to-all mode can be executed at same time, the ring mode only increase $n/m$ and the all-to-all mode inside blocks only increase $m(m-1)$ in circuit depth. The total circuit depth is $\left(\frac{n}{m} + m^2 - m + 4\right)L$. Similarly, we can calculate the complexity of Circuit 5 and Circuit 7.

Table 1: Cost estimates for 3 circuits, $n$ is the number of qubits, $L$ is the number of circuit layers, and $m$ is the number of qubits in a block defined in BR topology, $1 < m < n$.

| Circuit ID | Number of parameters | Number of 2-qubit gates | Circuit depth |
|---|---|---|---|
| 5 | $\left(3n + \frac{n}{gcd(n,3)}\right)L$ | $\left(n + \frac{n}{gcd(n,3)}\right)L$ | $\left(n + 2 + \frac{n}{gcd(n,3)}\right)L$ |
| 7 | $(n^2 + 3n)L$ | $(n^2 - n)L$ | $(n^2 - n + 4)L$ |
| 9 | $(m+4)nL$ | $mnL$ | $\left(\frac{n}{m} + m^2 - m + 4\right)L$ |

In **Table** 1, the number of parameters, the number of 2-qubits gates and circuit depth of circuit 5 are $O(n)$. Although the cost of circuit 5 is not high, the performance can not catch up with the performance of circuit 9. In the meantime, the cost of circuit 7 is $O(n^2)$, compared with circuit 7, the number of parameters and 2-qubit gate in circuit 9 reduce from $O(n^2)$ to $O(mn)$ without reducing performance. As long as m is chosen between 1 and n, the cost of circuit 9 always less than circuit 7.

## Conclusion

In this paper, we proposed a novel topology called BR topology for constructing PQCs. Experiments have proved that our BR topology have great expressibility and entangling capability. Compared with all-to-all topology, the BR topology saves the number of parameters, the number of 2-qubit gates, and greatly reduces the depth of PQCs.

The choice of $m$ has great influence on the performance of circuit, $1 < m < n$. In BR topology, the BR topology is close to ring topology when the value of $m$ tends to 1; when the value of $m$ tends to $n$, our topology is more similar to all-to-all topology. For different algorithms, researchers can select different values of $m$ to construct the circuit.

# Acknowledgements